\documentclass[preprint,12pt]{elsarticle}

\usepackage{graphicx}
\usepackage{amsmath,amssymb}
\usepackage{hyperref}
\usepackage{subfig}
\usepackage{caption}
\allowdisplaybreaks

\journal{Nuclear Instruments and Methods in Physics Research Section A}

\begin{document}

\begin{frontmatter}

\title{A Helical-Deflector-Based Radio-Frequency Spiral Scanning System for keV Energy Electrons}

\author[inst1]{Simon Zhamkochyan}
\author[inst1]{Vanik Kakoyan\texorpdfstring{\corref{cor1}}{*}}
\ead{kakoyan@yerphi.am}
\author[inst1]{Vardan Bardakhchyan}
\author[inst1]{Sergey Abrahamyan}
\author[inst1]{Amur Margaryan}
\author[inst1]{Aram Kakoyan}
\author[inst1]{Hasmik Rostomyan}
\author[inst1]{Anna Safaryan}
\author[inst1]{Gagik Sughyan}
\author[inst1]{Hayk Gevorgyan}
\author[inst1]{Artashes Papyan}
\author[inst1]{Martin Pinamyan}
\author[inst2]{Mikayel Ivanyan}
\author[inst3]{Satoshi N. Nakamura}
\author[inst4]{John Annand}
\author[inst4]{Kenneth Livingston}
\author[inst4]{Rachel Montgomery}
\author[inst5]{Patrick Achenbach}
\author[inst5]{Josef Pochodzalla}
\author[inst6]{Dimiter L. Balabanski}
\author[inst7]{Ani Aprahamian}
\author[inst8]{Viatcheslav Sharyy}
\author[inst8]{Dominique Yvon}
\author[inst1]{Hayk Elbakyan}

\cortext[cor1]{Corresponding author}

\affiliation[inst1]{organization={A.I. Alikhanyan National Science Laboratory (Yerevan Physics Institute)}, city={Yerevan}, country={Armenia}}
\affiliation[inst2]{organization={CANDLE Synchrotron Research Institute}, city={Yerevan}, country={Armenia}}
\affiliation[inst3]{organization={Department of Physics, Graduate School of Science, the University of Tokyo}, city={Tokyo}, country={Japan}}
\affiliation[inst4]{organization={School of Physics and Astronomy, University of Glasgow}, city={Glasgow}, postcode={G12 8QQ}, country={UK}}
\affiliation[inst5]{organization={Institut für Kernphysik, Johannes Gutenberg-Universität Mainz}, city={Mainz}, country={Germany}}
\affiliation[inst6]{organization={Extreme Light Infrastructure- Nuclear Physics (ELI-NP)}, city={Bucharest-Magurele}, country={Romania}}
\affiliation[inst7]{organization={Department of Physics and Astronomy, University of Notre Dame}, city={Notre Dame, IN 46556}, country={USA}}
\affiliation[inst8]{organization={Département de Physique des Particules Centre de Saclay}}

\begin{abstract}
We present the design, modeling, and experimental validation of a radio-frequency based time-to-position conversion system for keV electrons incorporating a helical deflector operating in the 400-1000\,MHz range. The device performs circular deflection of the electrons when driven by a single RF frequency and enables spiral scanning when two phase-locked RF voltages with slightly different frequencies are applied. The superposition of the two phase-locked RF voltages produces an amplitude-beating field whose slowly varying envelope modulates the deflection radius, transforming the circular scan into a controlled spiral on the detector plane. A detailed theoretical model describing the electron dynamics under two phase-locked RF voltages with different frequencies was derived, yielding analytical expressions for the transverse velocity and radius-vector components at the deflector exit. The experimental studies demonstrated good agreement with the model predictions. Spiral scanning will allow measurements with picosecond resolution in a temporal dynamic range 1-2 orders of magnitude larger than the period of the circular scanning.
\end{abstract}


\end{frontmatter}

\section{Introduction}
High-precision timing is indispensable in many disciplines, including particle and nuclear physics, astrophysics, ultrafast chemistry, biomedical imaging, and materials science. Here we report a development of a radio-frequency (RF) timer for keV-energy electrons based on a helical RF deflector operating in the 400-1000\,MHz range \cite{Gevorgian2015,Margaryan2022}. The deflector converts the arrival time of each electron into a position along a circular locus by performing continuous RF sweeps, enabling direct time-to-position mapping with low dead time. This concept is closely related to the RF photomultiplier tube (RFPMT) \cite{Margaryan2006}, which combines RF deflection with fast electron detection. Detection is performed using a microchannel plate (MCP) assembly coupled to a delay-line (DL) anode, which records the spatial distribution of impacts corresponding to the temporal structure of the incident electrons. Using an RF-synchronized femtosecond laser, a timing resolution of less than 10\,ps has been demonstrated \cite{Margaryan2022}, limited mainly by the prototype geometry and operating frequency. The intrinsic dead time of the RFPMT is determined by the readout system; for the implemented DL anode it is around 40\,ns. The duration of one circular sweep corresponds to the inverse of the RF frequency: 2\,ns at 500\,MHz and 0.1\,ns at 10\,GHz. To extend the temporal dynamic range while maintaining ultrafast precision, we introduce a spiral-scanning technique employing two slightly different, phase-locked RF frequencies $\omega_1$ and $\omega_2$ \cite{Gurov2007,vanRens2018,Elbakyan2016}. The resulting pulsed image forms a slowly rotating spiral trajectory, increasing the duration of a single scan by several tens of times. The following sections present the theoretical background of the RF spiral-scanning system (Section~\ref{sec:theory}), experimental setup and results (Section~\ref{sec:experiment}), and concluding remarks (Section~\ref{sec:conclusion}).

\section{Theoretical Concept of the Spiral Scanning}
\label{sec:theory}
\begin{figure}[ht]
\centering
\includegraphics[width=\linewidth]{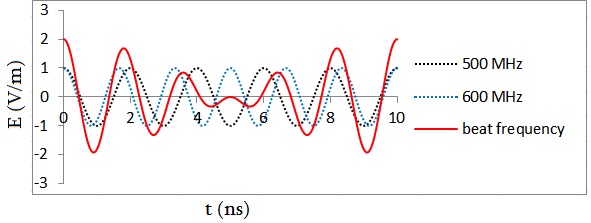}
\caption{Beat phenomenon for frequencies of 500\,MHz and 600\,MHz.}
\label{fig:fig1_beatings}
\end{figure}

The spiral-scanning principle is based on the beat phenomenon that arises from the superposition of two sinusoidal voltages with sufficiently close frequencies, $f_1$ and $f_2$. The resulting electric field can be expressed as
\begin{equation}
\begin{aligned}
E(t)=E_1(t)+E_2(t)=E_d\cos(2\pi f_1 t)+E_d\cos(2\pi f_2 t)=\\
2E_d\cos\!\left(2\pi\frac{f_1-f_2}{2}t\right)\cos\!\left(2\pi\frac{f_1+f_2}{2}t\right).
\end{aligned}
\tag{1}
\end{equation}
Here, $f_b=|f_1-f_2|$ is the beat frequency, which determines the rate at which the amplitude envelope of the combined signal oscillates. For example, when $f_1=500$\,MHz, $f_2=600$\,MHz, and $E_d=1$\,V/m, a distinct beat pattern with $T_b=\dfrac{1}{f_b}=10$\,ns is observed, as shown in Fig.~\ref{fig:fig1_beatings}. For the case, when $f_1=500$\,MHz and $f_2=505$\,MHz the $T_b=200$\,ns.

\begin{figure}[t]
\centering
\includegraphics[width=0.9\linewidth]{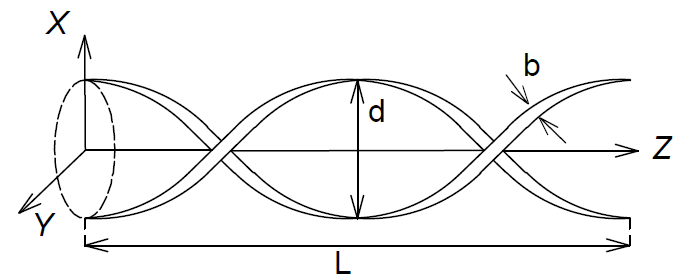}
\caption{Schematic of the deflector.}
\label{fig:fig2_deflector}
\end{figure}

In the previous study \cite{Gevorgian2015}, a theoretical model of a helical deflector operating in the single frequency range of 400-1000\,MHz was presented, along with experimental results. This deflector (Fig.~\ref{fig:fig2_deflector}), composed of two helical electrodes of axial length L and width b separated by a distance d. This type of RF deflector was proposed by Shamaev in the 1960-s to avoid transit-time effects at high frequencies (see details in \cite{Zhigarev1972}). 

Electrons enter the deflector along the Z-axis with a constant non-relativistic velocity $V_z$. For circular or elliptical scanning, the deflector is driven by an RF voltage of angular frequency $\omega_1$. As the electrons pass through the deflector, their transverse velocity components $V_{x1}$ and $V_{y1}$ at the end of the deflector are given by \cite{Gevorgian2015}:
\begin{equation}
V_{x1}=-\frac{eE_{d1}\tau}{2m_e}\left[
\frac{\sin x_2}{x_2}\sin(x_2+\varphi_1)-\frac{\sin x_1}{x_1}\sin(x_1-\varphi_1)
\right],
\tag{2}
\end{equation}
\begin{equation}
V_{y1}=-\frac{eE_{d1}\tau}{2m_e}\left[
\frac{\sin x_1}{x_1}\cos(x_1-\varphi_1)-\frac{\sin x_2}{x_2}\cos(x_2+\varphi_1)
\right].
\tag{3}
\end{equation}
Here $e$ and $m_e$ are the charge and mass of the electron, respectively, $E_{d1}$ and $\varphi_1=\omega_1 t+\varphi_{01}$ are the amplitude and phase of the RF voltage, respectively, $\varphi_{01}$ is the initial phase of the RF voltage at $t=0$, and $\tau=L/V_z$ is the transit time of the electron through the deflector of length L. The parameters $x_1$ and $x_2$ are defined as
\begin{equation}
x_1=\frac{\omega_c-\omega_1}{2}\tau,\qquad x_2=\frac{\omega_c+\omega_1}{2}\tau,
\tag{4}
\end{equation}
where $\omega_c=(2\pi V_z)/\lambda_c$ is a parameter characterizing the deflector, with $\lambda_c$ the winding pitch of the helical electrodes. We have circular (elliptical) scanning in the case of $\omega_1=\omega_c$ ($\omega_1\neq\omega_c$). The length of the deflecting system (L) is chosen to be proportional to $\lambda_c$/4~\cite{Gevorgian2015}. Subsequent calculations are made for $L=\lambda_c$.

When two RF voltages with different angular frequencies $\omega_1$ and $\omega_2$ are applied to the deflector, the electron motion becomes the superposition of two independent circular (elliptical) motions. The transverse velocity components at the end of the deflector in this case can be written as

\begin{equation}
\begin{aligned}
V_x=V_{x1}+V_{x2}=-\frac{eE_{d1}\tau}{2m_e}\left[\frac{\sin x_2}{x_2}\sin(x_2+\varphi_1)-\frac{\sin x_1}{x_1}\sin(x_1-\varphi_1)\right]\\
-\frac{eE_{d2}\tau}{2m_e}\left[\frac{\sin x_2'}{x_2'}\sin(x_2'+\varphi_2)-\frac{\sin x_1'}{x_1'}\sin(x_1'-\varphi_2)\right],    
\end{aligned}
\tag{5}
\end{equation}
\begin{equation}
\begin{aligned}
V_y=V_{y1}+V_{y2}=-\frac{eE_{d1}\tau}{2m_e}\left[\frac{\sin x_1}{x_1}\cos(x_1-\varphi_1)-\frac{\sin x_2}{x_2}\cos(x_2+\varphi_1)\right]\\
-\frac{eE_{d2}\tau}{2m_e}\left[\frac{\sin x_1'}{x_1'}\cos(x_1'-\varphi_2)-\frac{\sin x_2'}{x_2'}\cos(x_2'+\varphi_2)
\right],
\end{aligned}
\tag{6}
\end{equation}
where $E_{d2}$ and $\varphi_2$ are the amplitude and phase of the second RF voltage, respectively, and the parameters $x_1'$ and $x_2'$ are related as
\begin{equation}
x_1'=\frac{\omega_c-\omega_2}{2}\tau,\qquad x_2'=\frac{\omega_c+\omega_2}{2}\tau.
\tag{7}
\end{equation}
Introducing the parameters
\begin{equation}
k_1=\frac{\omega_1}{\omega_c},\qquad k_2=\frac{\omega_2}{\omega_c}
\tag{8}
\end{equation}
and considering equations (4)-(8), we obtain:
\begin{equation}
\begin{aligned}
x_1&=\frac{\omega_c-\omega_1}{2}\tau=(1-k_1)\frac{\omega_c}{2}\tau=(1-k_1)\pi,\\
x_2&=\frac{\omega_c+\omega_1}{2}\tau=(1+k_1)\frac{\omega_c}{2}\tau=(1+k_1)\pi,\\
x_1'&=\frac{\omega_c-\omega_2}{2}\tau=(1-k_2)\frac{\omega_c}{2}\tau=(1-k_2)\pi,\\
x_2'&=\frac{\omega_c+\omega_2}{2}\tau=(1+k_2)\frac{\omega_c}{2}\tau=(1+k_2)\pi.
\end{aligned}
\tag{9}
\end{equation}
The phase of the second voltage is expressed as $\varphi_2=\omega_2 t+\varphi_{02}$, where $\varphi_{02}$ is the initial phase of the second voltage at $t=0$. Let’s assume that the initial phase of first RF voltage is $\varphi_{01}=0$ and that of the second RF voltage is $\varphi_{02}=\Delta\varphi$. Considering equation (8), we get:
\begin{equation}
\varphi_2=\frac{k_2}{k_1}\varphi_1+\Delta\varphi.
\tag{10}
\end{equation}
Substituting equations (9) and (10) into expressions (5) and (6), and replacing $\varphi_1$ with $\varphi$, we obtain:

\begin{equation}
\begin{aligned}
\frac{V_x}{V_0}={}&-\left[\frac{\sin(1+k_1)\pi}{(1+k_1)\pi}
\sin\!\left((1+k_1)\pi+\varphi\right)\right.\\
&\left.\qquad
-\frac{\sin(1-k_1)\pi}{(1-k_1)\pi}
\sin\!\left((1-k_1)\pi-\varphi\right)\right]\\
&-a\left[\frac{\sin(1+k_2)\pi}{(1+k_2)\pi}
\sin\!\left((1+k_2)\pi+\frac{k_2}{k_1}\varphi+\Delta\varphi\right)\right.\\
&\left.\qquad
-\frac{\sin(1-k_2)\pi}{(1-k_2)\pi}
\sin\!\left((1-k_2)\pi-\frac{k_2}{k_1}\varphi-\Delta\varphi\right)\right],
\end{aligned}
\tag{11}
\end{equation}
\begin{equation}
\begin{aligned}
\frac{V_y}{V_0}={}&-\left[\frac{\sin(1-k_1)\pi}{(1-k_1)\pi}
\cos\!\left((1-k_1)\pi-\varphi\right)\right.\\
&\left.\qquad
-\frac{\sin(1+k_1)\pi}{(1+k_1)\pi}
\cos\!\left((1+k_1)\pi+\varphi\right)\right]\\
&-a\left[\frac{\sin(1-k_2)\pi}{(1-k_2)\pi}
\cos\!\left((1-k_2)\pi-\frac{k_2}{k_1}\varphi-\Delta\varphi\right)\right.\\
&\left.\qquad
-\frac{\sin(1+k_2)\pi}{(1+k_2)\pi}
\cos\!\left((1+k_2)\pi+\frac{k_2}{k_1}\varphi+\Delta\varphi\right)\right],
\end{aligned}
\tag{12}
\end{equation}
where $\varphi$ is the phase of the first RF, $a=E_{d2}/E_{d1}$ and $V_0=eE_{d1}\tau/2m_e$ are constant parameters.

To find the transverse components of the electron's radius vector on the plane at the end of the deflector, we integrate equations (11) and (12) with respect to the time (t) of the RF voltage over the interval from $t'$ to $t'+\tau$, where $t'$ and $t'+\tau$ are the times at which the electron enters and leaves the deflector, respectively.

\begin{equation}
\begin{aligned}
r_x = -V_0 \int_{t'}^{t'+\tau} \Bigg[ &
\frac{\sin\!\big((1+k_1)\pi\big)}{(1+k_1)\pi}\,
\sin\!\big((1+k_1)\pi+\omega_1 t\big) \\
& -
\frac{\sin\!\big((1-k_1)\pi\big)}{(1-k_1)\pi}\,
\sin\!\big((1-k_1)\pi-\omega_1 t\big)
\Bigg] \\
& - a \Bigg[ \frac{\sin\!\big((1+k_2)\pi\big)}{(1+k_2)\pi}\,
\sin\!\big((1+k_2)\pi+\frac{k_2}{k_1}\omega_1 t+\Delta\varphi\big) \\
& - \frac{\sin\!\big((1-k_2)\pi\big)}{(1-k_2)\pi}\,
\sin\!\big((1-k_2)\pi-\frac{k_2}{k_1}\omega_1 t-\Delta\varphi\big)
\Bigg] \, dt,
\end{aligned}
\tag{13}
\end{equation}

\begin{equation}
\begin{aligned}
r_y = -V_0 \int_{t'}^{t'+\tau} \Bigg[ &
\frac{\sin\!\big((1-k_1)\pi\big)}{(1-k_1)\pi}\,
\cos\!\big((1-k_1)\pi-\omega_1 t\big) \\
& -
\frac{\sin\!\big((1+k_1)\pi\big)}{(1+k_1)\pi}\,
\cos\!\big((1+k_1)\pi+\omega_1 t\big)
\Bigg] \\
& - a \Bigg[\frac{\sin\!\big((1-k_2)\pi\big)}{(1-k_2)\pi}\,
\cos\!\big((1-k_2)\pi-\frac{k_2}{k_1}\omega_1 t-\Delta\varphi\big) \\
& - \frac{\sin\!\big((1+k_2)\pi\big)}{(1+k_2)\pi}\,
\cos\!\big((1+k_2)\pi+\frac{k_2}{k_1}\omega_1 t+\Delta\varphi\big)
\Bigg] \, dt.
\end{aligned}
\tag{14}
\end{equation}

After integration we get:

\begin{equation}
\begin{aligned}
\frac{r_x}{r_0}=&
\frac{\sin(1+k_1)\pi}{(1+k_1)\pi}
\left[\cos\!\left((1+k_1)\pi+\varphi+2\pi k_1\right)
-\cos\!\left((1+k_1)\pi+\varphi\right)\right]\\
&+\frac{\sin(1-k_1)\pi}{(1-k_1)\pi}
\left[\cos\!\left((1-k_1)\pi-\varphi-2\pi k_1\right)
-\cos\!\left((1-k_1)\pi-\varphi\right)\right]\\
&+a\frac{k_1}{k_2}\frac{\sin(1+k_2)\pi}{(1+k_2)\pi}
\left[\cos\!\left((1+k_2)\pi+\frac{k_2}{k_1}\varphi+2\pi k_2+\Delta\varphi\right)\right.\\
&\left.\qquad
-\cos\!\left((1+k_2)\pi+\frac{k_2}{k_1}\varphi+\Delta\varphi\right)\right]\\
&+a\frac{k_1}{k_2}\frac{\sin(1-k_2)\pi}{(1-k_2)\pi}
\left[\cos\!\left((1-k_2)\pi-\frac{k_2}{k_1}\varphi-2\pi k_2-\Delta\varphi\right)\right.\\
&\left.\qquad
-\cos\!\left((1-k_2)\pi-\frac{k_2}{k_1}\varphi-\Delta\varphi\right)\right],
\end{aligned}
\tag{15}
\end{equation}

\begin{equation}
\begin{aligned}
\frac{r_y}{r_0}=&
\frac{\sin(1+k_1)\pi}{(1+k_1)\pi}
\left[\sin\!\left((1+k_1)\pi+\varphi+2\pi k_1\right)
-\sin\!\left((1+k_1)\pi+\varphi\right)\right]\\
&+\frac{\sin(1-k_1)\pi}{(1-k_1)\pi}
\left[\sin\!\left((1-k_1)\pi-\varphi-2\pi k_1\right)
-\sin\!\left((1-k_1)\pi-\varphi\right)\right]\\
&+a\frac{k_1}{k_2}\frac{\sin(1+k_2)\pi}{(1+k_2)\pi}
\left[\sin\!\left((1+k_2)\pi+\frac{k_2}{k_1}\varphi+2\pi k_2+\Delta\varphi\right)\right.\\
&\left.\qquad
-\sin\!\left((1+k_2)\pi+\frac{k_2}{k_1}\varphi+\Delta\varphi\right)\right]\\
&+a\frac{k_1}{k_2}\frac{\sin(1-k_2)\pi}{(1-k_2)\pi}
\left[\sin\!\left((1-k_2)\pi-\frac{k_2}{k_1}\varphi-2\pi k_2-\Delta\varphi\right)\right.\\
&\left.\qquad
-\sin\!\left((1-k_2)\pi-\frac{k_2}{k_1}\varphi-\Delta\varphi\right)\right],
\end{aligned}
\tag{16}
\end{equation}

where $\varphi$ is the phase of the first RF at the moment $t'$ and $r_0=\frac{V_0}{\omega_1}=eE_{d1}\tau^2/4k_1\pi m_e$.

As stated above, the theoretical analysis provides explicit expressions for the transverse components of the electron’s radius vector, which determined by following parameters of the deflector:
\begin{enumerate}
\item $a$ - the amplitude ratio of the two RF voltages;
\item $\Delta\varphi$ - the initial phase difference of the two RF voltages;
\item $k_1$ and $k_2$ which are defined by the $f_1$, $f_2$ frequencies and the $\omega_c$ parameter of the helical deflector.
\end{enumerate}

\section{Experimental Studies}
\label{sec:experiment}
\begin{figure}[ht]
\centering
\includegraphics[width=\linewidth]{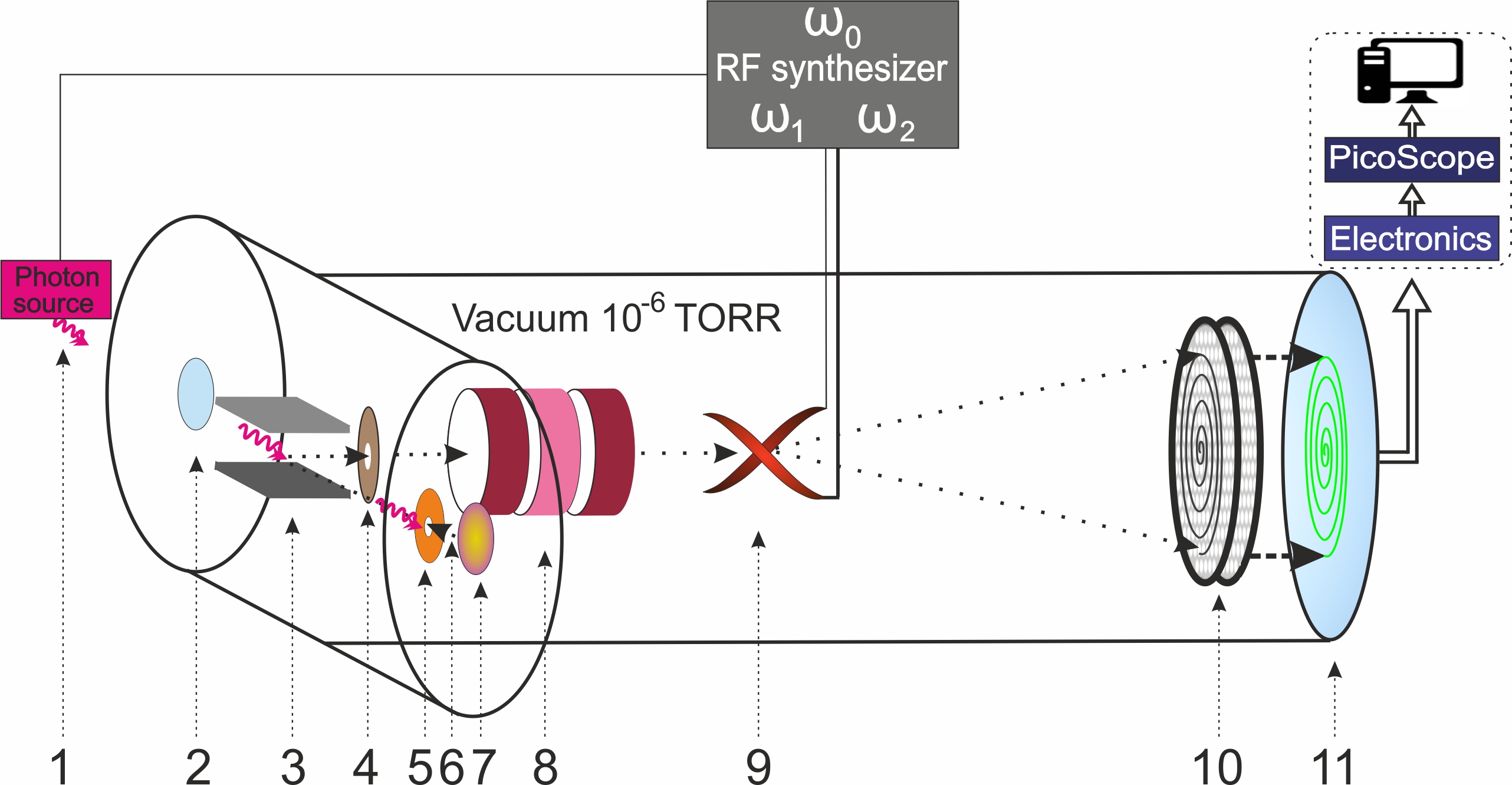}
\caption{Schematic of the experimental setup. 1 - UV photons; 2 - quartz window; 3 - magnet; 4 - collimator; 5 - accelerating electrode; 6 - photoelectron; 7 - cathode; 8 - electrostatic lens; 9 - RF deflector; 10 - MCP detector; 11 - delay line anode.}
\label{fig:fig3_exp_setup}
\end{figure}

Experiments were performed using the setup schematically illustrated in Fig.~\ref{fig:fig3_exp_setup}. The system consists of several components: a UV photon source, an RF timer, an RF source, associated electronics, a data acquisition (DAQ) system, power supplies, and a vacuum chamber. The components of the RF timer \cite{Margaryan2022} are assembled in a single tube and operated under a vacuum of $10^{-6}$ Torr or better.

An ultra-low-noise femtosecond laser module (ORIGAMI-05LP, NKT Photonics) was used as the photon source. The laser provides pulses with a duration of 166~fs at a wavelength of 515~nm, with a repetition rate of 40\,MHz and an average power of 25~mW. A beta-barium borate~(BBO) nonlinear crystal frequency-doubles the laser output from 515~nm to 257.5~nm.
The resulting photons~(1) pass through a quartz window and are directed onto a gold photocathode~(7). The emitted low-energy electron beam~(6) is accelerated between the cathode and the accelerating electrode~(5) by a potential difference of 2.5\,kV. The accelerated electrons are deflected by 90$^\circ$ using a permanent magnet~(3) and collimated by an aperture of 0.7\,mm~(4). They are then focused by an electrostatic lens~(8) and, after passing through a spiral RF deflector~(9), which can be synchronized with the laser's repetition frequency, are directed onto a position-sensitive detector~(PSD)~\cite{delay_line}.

The accelerating electrode is a copper disk with a central aperture of 2.0\,mm diameter, positioned 2\,mm from the cathode. The PSD consists of a double-chevron microchannel plate (MCP) assembly~(10) coupled to a delay-line~(DL) anode~(11). The active diameter of the MCP is 25\,mm. Within the MCPs, electrons are multiplied by a factor of approximately~10$^6$ and directed onto the DL anode, generating position-dependent signals with nanosecond rise times.
The RF deflector is driven by an RF synthesizer and is located approximately 120\,mm upstream of the MCP. The signal pairs generated at the outputs of the delay-line anode are extracted through vacuum feedthroughs, amplified using low-noise amplifiers, and digitized with a PicoScope device~\cite{PicoTech}. The electron impact positions on the PSD are reconstructed from the differences in the signal arrival times.

\begin{figure}[ht]
\centering
\subfloat[(a)]{\includegraphics[width=0.32\linewidth]{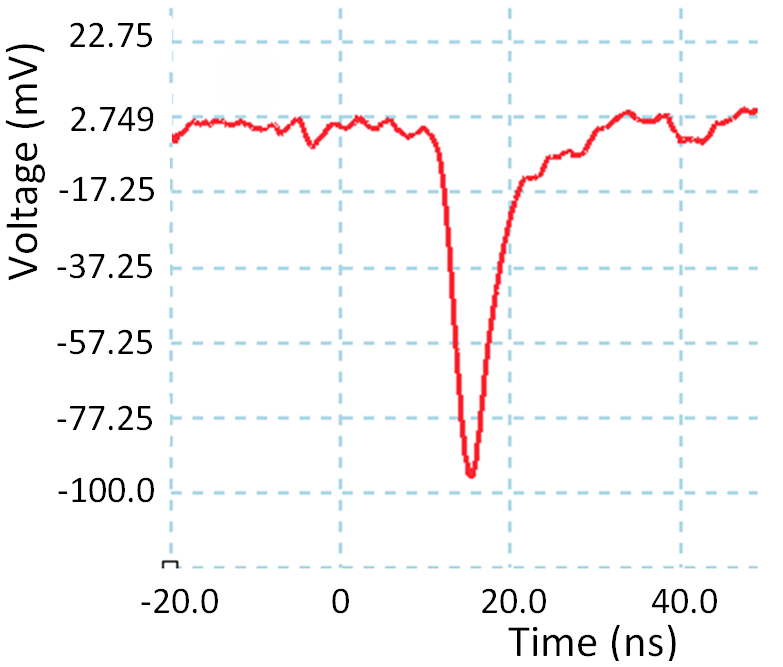}\label{fig:fig4a_signal}}\hfill
\subfloat[(b)]{\includegraphics[width=0.32\linewidth]{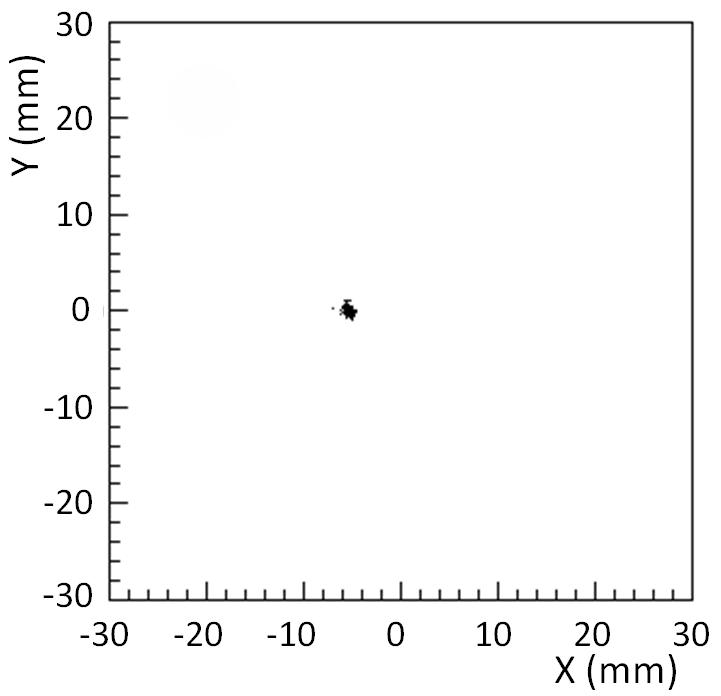}\label{fig:fig4b_rf_off}}\hfill
\subfloat[(c)]{\includegraphics[width=0.32\linewidth]{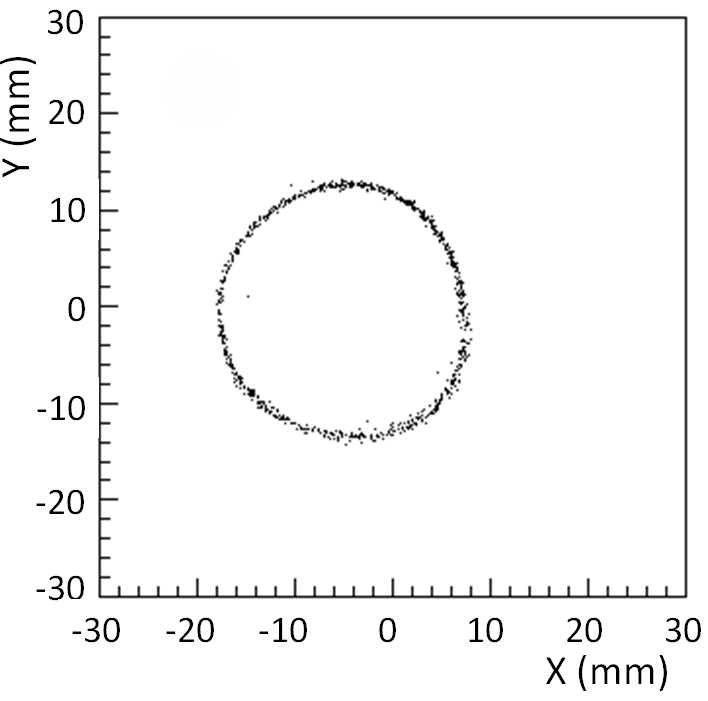}\label{fig:fig4c_rf_on}}
\caption{(a) typical amplified signals from the DL anode; (b) 2D image of the focused electrons (RF is OFF); (c) 2D image of the scanned electrons (500\,MHz RF is ON).}
\end{figure}

Figure~\ref{fig:fig4a_signal} shows the typical signal from the DL anode after amplification. When no RF is applied to the deflector, the image of the focused electrons on the anode is a dot (Fig.~\ref{fig:fig4b_rf_off}), while an ellipse or circle is obtained (Fig.~\ref{fig:fig4c_rf_on}) when RF is applied. The X and Y distributions of the focused electrons has a gaussian distributions with $\sigma\sim$150\,$\mu$m.

\begin{figure}[ht]
\centering
\includegraphics[width=0.75\linewidth]{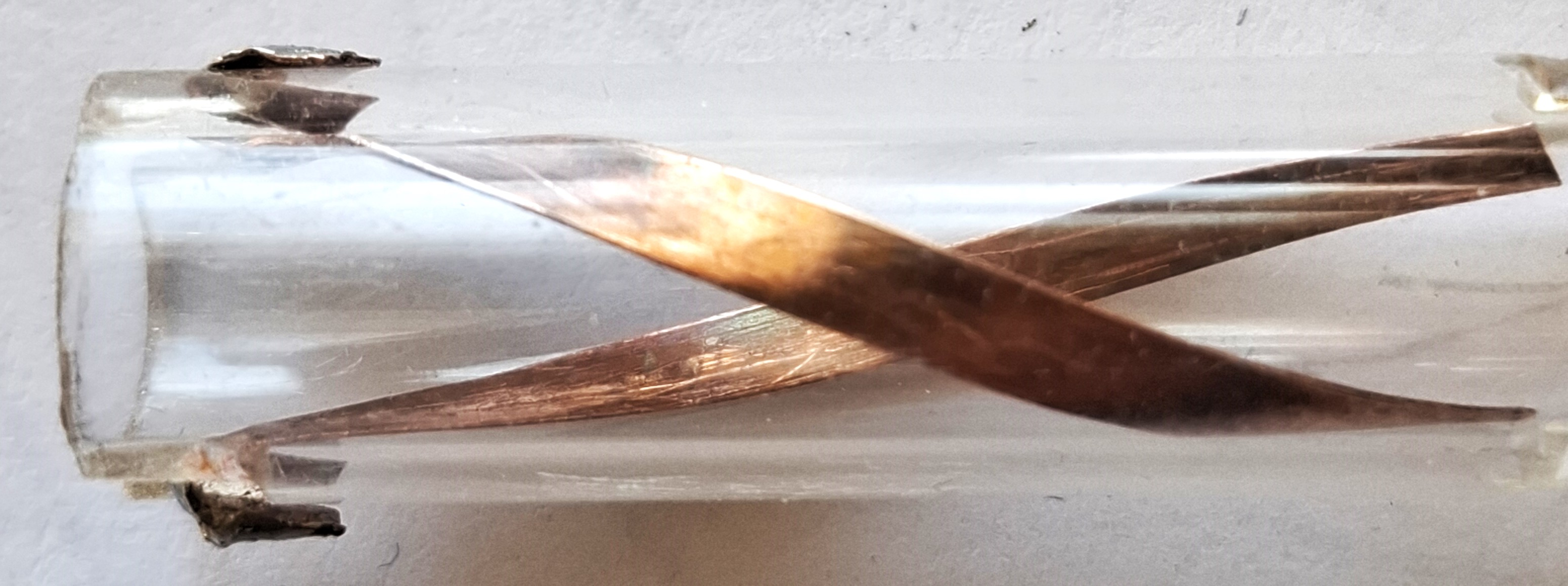}
\caption{Photograph of the half-period deflector.}
\label{fig:fig5_deflector_half_photo}
\end{figure}

Experimental studies were performed using two different helical RF deflectors (half- and full-period), each with dimensions optimized for a frequency of 500\,MHz. A photograph of the first, half-period deflector is shown in Fig.~\ref{fig:fig5_deflector_half_photo}. The deflector is 30\,mm long and is made up of two thin copper electrodes, each 3\,mm in width. These electrodes are positioned inside a quartz tube with an inner diameter of 7.6\,mm, arranged diametrically opposite each other. 

\begin{figure}[ht]
\centering
\subfloat[(a)]{\includegraphics[width=0.23\linewidth]{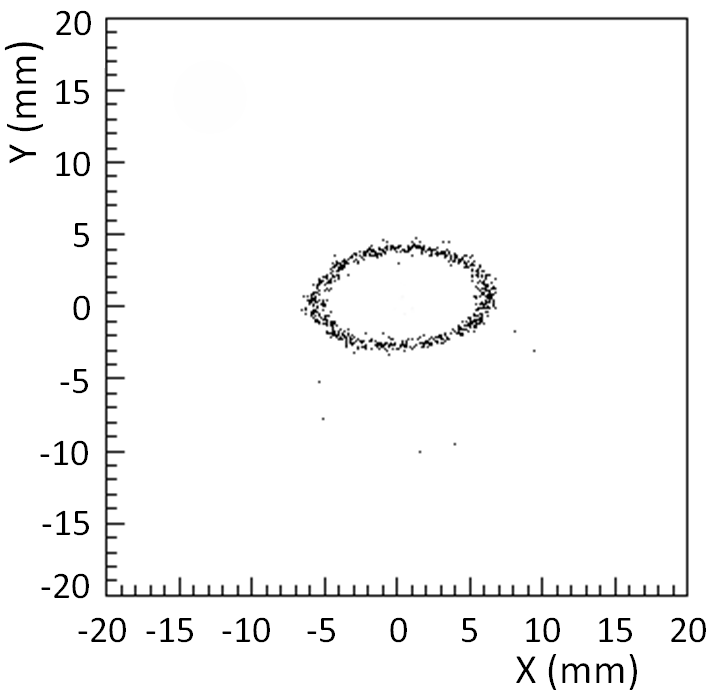}\label{fig:fig6a_halfdefl_500}}\hfill
\subfloat[(b)]{\includegraphics[width=0.23\linewidth]{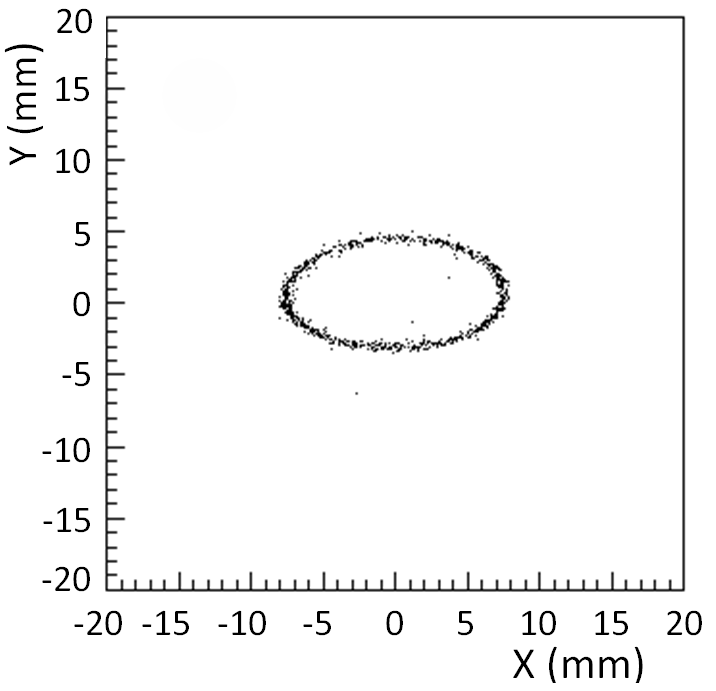}\label{fig:fig6b_halfdefl_600}}\hfill
\subfloat[(c)]{\includegraphics[width=0.23\linewidth]{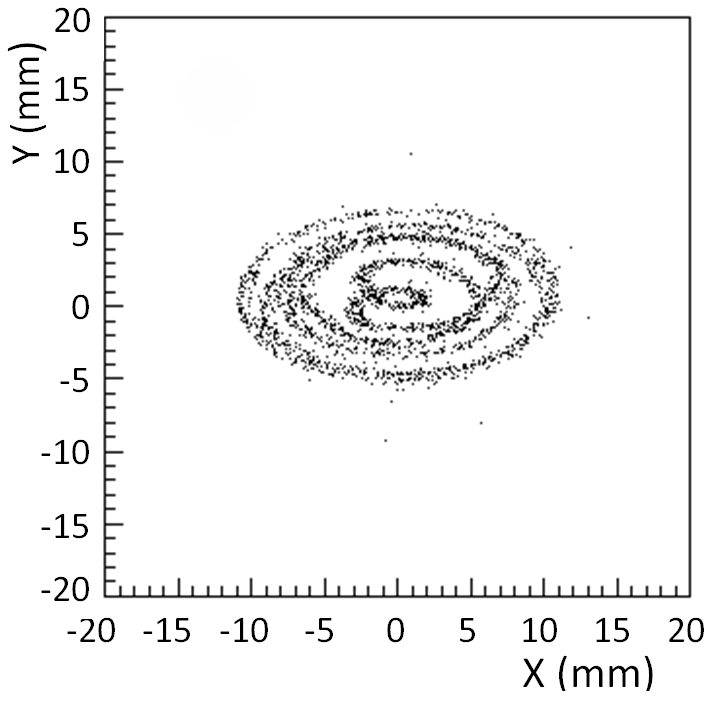}\label{fig:fig6c_combined}}\hfill
\subfloat[(d)]{\includegraphics[width=0.23\linewidth]{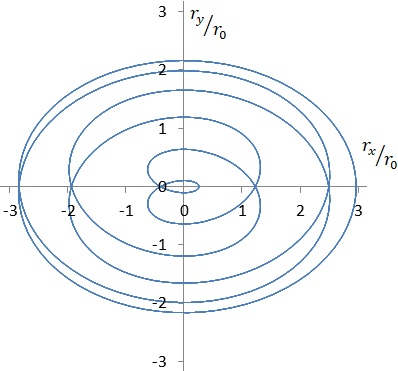}\label{fig:fig6d_theory}}
\caption{Half-period deflector. Two-dimensional images of scanned electrons: (a) $f_1=500$\,MHz, (b) $f_2=600$\,MHz, (c) the combined frequencies $f_1+f_2$ and (d) the corresponding theoretical prediction with $f_c=405$\,MHz, $a=1.23$ and $\Delta\varphi=\pi/2$. The temporal dynamic range is $T_b=10$\,ns.}
\label{fig:fig6_halfdefl}
\end{figure}

Subsequent measurements were performed using two different phase-locked RF frequencies. When only one RF signal was applied to the deflector the ellipse or circle was obtained on the anode. When two RF signals were applied, then a spiral was obtained. Figure~\ref{fig:fig6_halfdefl} shows an example of experimental and theoretical results for two different RF signal parameters (see Section~\ref{sec:theory}). It should be noted that, in the present experimental conditions, the $\Delta\varphi$ is neither controlled nor independently measured. For this reason, a direct point-by-point quantitative comparison between the calculated and measured traces is not uniquely defined. Instead, the comparison is performed at the level of the global image morphology.
Specifically, we evaluate the agreement in terms of the characteristic geometrical features of the spiral patterns, including their topology, orientation, and the relative spacing of successive turns. As seen in Fig.~\ref{fig:fig6_halfdefl}, the theoretical model reproduces the main features of the observed patterns, for physically reasonable choices of the phase offsets. The temporal dynamic range of a single circular spectrum, which lasts about 2\,ns at frequencies $f_1=500$\,MHz or $f_2=600$\,MHz, increases to $T_b=10$\,ns when two RF signals are applied together.

\begin{figure}[ht]
\centering
\includegraphics[width=0.9\linewidth]{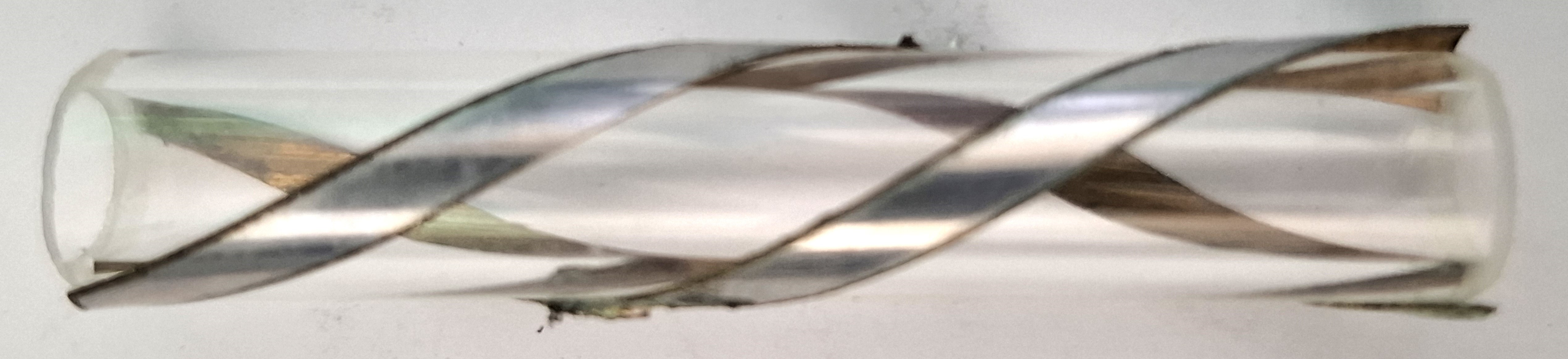}
\caption{Photograph of the full period deflector.}
\label{fig:fig7_deflector_full_photo}
\end{figure}

 Consequently, $f_c$ serves as an effective normalization parameter between the theoretical and experimental results.

The second, full-period deflector was placed on the outer surface of a 12\,mm diameter quartz tube (see Fig.~\ref{fig:fig7_deflector_full_photo}). It has a length of 60\,mm and the width of the copper electrodes is 5\,mm.

\begin{figure}[ht]
\centering
\subfloat[(a)]{\includegraphics[width=0.23\linewidth]{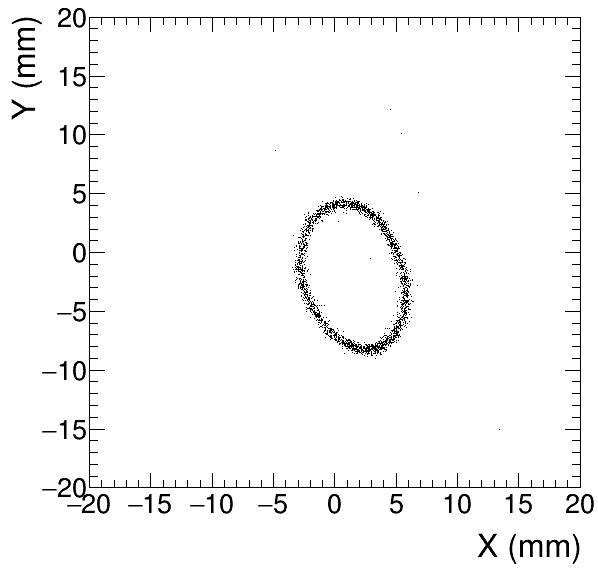}}\hfill
\subfloat[(b)]{\includegraphics[width=0.23\linewidth]{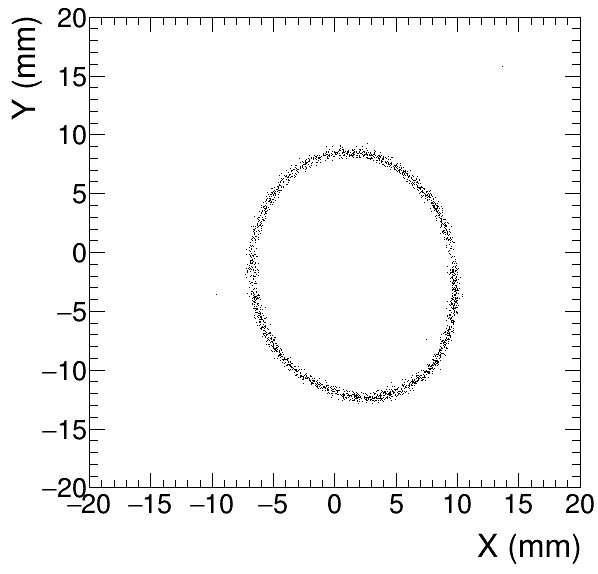}}\hfill
\subfloat[(c)]{\includegraphics[width=0.23\linewidth]{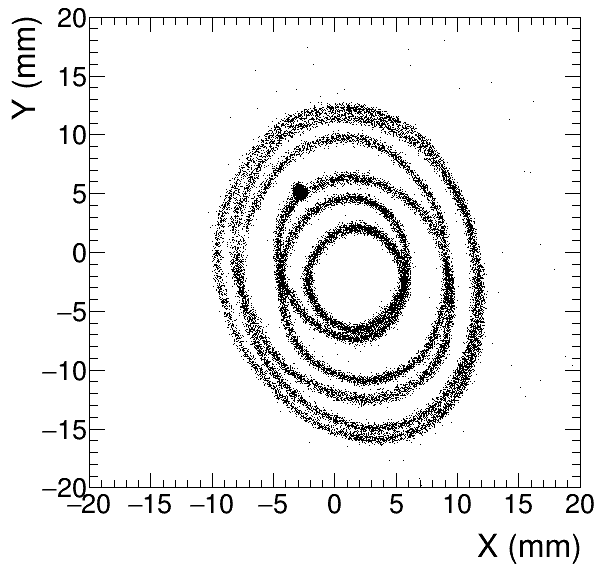}\label{fig:fig8c_combined}}\hfill
\subfloat[(d)]{\includegraphics[width=0.23\linewidth]{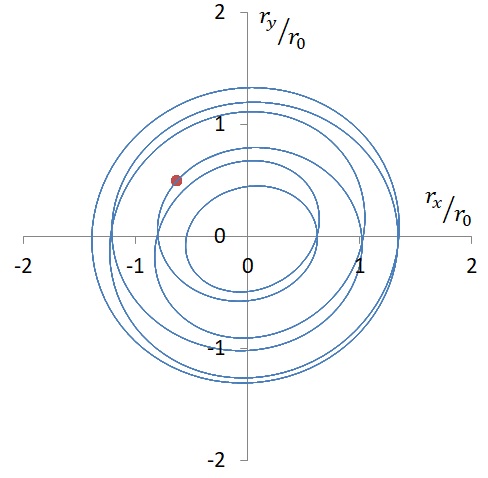}\label{fig:fig8d_theory}}
\caption{Full-period deflector. Two-dimensional images of scanned electrons: (a)$f_1=400$\,MHz, (b) $f_2=480$\,MHz, (c) the combined frequencies $f_1+f_2$ (spiral image), where the localized spot on the spiral corresponds to the phase distribution of RF-synchronized photoelectrons and (d) the corresponding theoretical prediction with $f_c=430$\,MHz, $a=1.6$ and $\Delta\varphi=\pi/4$. The temporal dynamic range is $T_b=12.5$\,ns.}
\label{fig:fig8_fulldefl}
\end{figure}

An example of the theoretical and experimental results for this deflector is shown in Fig.~\ref{fig:fig8_fulldefl}. The spot, shown superimposed on the spiral image in Fig.~\ref{fig:fig8c_combined}, was obtained in RF-synchronized mode, where the 40\,MHz laser repetition rate was phase-locked to the RF signals at 400 and 480\,MHz driving the deflector. The spot in the theoretical pattern (Fig.~\ref{fig:fig8d_theory}) represents the electrons coordinates checked with 25\,ns period (40\,MHz repetition rate). The spot, shown superimposed on the spiral image in Fig.~\ref{fig:fig8c_combined}, was obtained in RF-synchronized mode, where the 40\,MHz laser repetition rate was phase-locked to the RF signals at 400 and 480\,MHz driving the deflector. The size of this spot, when converted into time units, represents the temporal resolution of the system. The observed spot size corresponds to a resolution of approximately 7\,ps, which can vary from 4 to 14\,ps depending on the effective radius. The resolution improves at a larger radii of the spiral due to the increased effective scanning radius.

These results demonstrate that a deflector with a given set of parameters can operate efficiently over a relatively wide range of frequencies.  While spiral scans are shown here for frequencies in the 400–600\,MHz range, circular scanning of the same deflectors has been experimentally verified up to 1000\,MHz~\cite{Gevorgian2015}. Consequently, by using a deflector with suitable parameters, it becomes possible to achieve spiral scanning with a specified beat period in the needed frequency region throughout the 400–1000\,MHz frequency range.

\section{Conclusion}
\label{sec:conclusion}
In this work, we report on the development and demonstration of a helical-deflector-based radio-frequency spiral scanning system capable of detecting keV energy electrons with a time resolution of a few picoseconds. A theoretical model describing the electron dynamics under two phase-locked RF voltages with different frequencies was derived, yielding analytical expressions for the transverse velocity and radius-vector components of electrons at the deflector exit. These expressions show that the resulting electron trace, and thus the time-to-position mapping, are fully determined by three fundamental parameters: the ratio of the RF amplitudes $a$, the initial phase offset $\Delta\varphi$, and the normalized frequencies $k_1$ and $k_2$. The model predicts the formation of well-defined spiral traces whose shape and temporal scale can be precisely controlled through these parameters. Experimental measurements performed with 2.5\,keV electrons confirm the theoretical predictions. Spiral scans obtained using RF frequencies in the 400-1000\,MHz range exhibit good agreement with simulations, validating both the helical-deflector concept and the two-frequency driving scheme. The spiral mode significantly extends the temporal dynamic range compared to conventional circular scanning, while retaining the underlying picosecond-level temporal resolution. 

The method provides high-rate detection using MCP delay-line readout and can alternatively employ pixelated detectors, such as Timepix4~\cite{Timepix}, which offer the potential for substantially reduced system dead time and detection rates approaching the THz regime~\cite{Margaryan2012}. This would allow detection of multiple electron events separated by only a few picoseconds within a single scan cycle. Thus, with further optimization, the approach offers a promising pathway toward very low dead-times and therefore, high-throughput timing devices for next-generation photodetectors.

\section*{Acknowledgments}
This work was supported by the Higher Education and Science Committee of the Republic of Armenia (Research Project: 23LCG-1C018) and the International Science and Technology Center (ISTC project AM-2803).

\section*{Declaration of Interest}

The authors declare that they have no known competing financial interests or personal relationships that could have appeared to influence the work reported in this paper.

\section*{CrediT authorship contribution statement}
Vanik Kakoyan: Management, Conceptualization, Investigation, Software, Resources. Simon Zhamkochyan: Software, Investigation, Resources, Writing – review \& editing. Vardan Bardakhchyan: Theory, Validation, Methodology. Sergey Abrahamyan: Software, Investigation, Writing – review \& editing. Amur Margaryan: Supervision, Conceptualization, Methodology, Writing – review \& editing. Aram kakoyan: Hardware development. Hasmik Rostomyan: Formal analysis, software. Anna Safaryan: Formal analysis, software. Gagik Sughyan: Hardware development, Electronics. Hayk Gevorgyan: Formal analysis. Artashes Papyan: Electronics. Martin Pinamyan: Hardware development. Mikayel Ivanyanan: Validation. John Annand: Investigation, Writing – review \& editing. Kenneth Livingston: Writing – review \& editing. Rachel Montgomery: Writing – review \& editing. Patrick Achenbach: Writing – review \& editing. Josef Pochodzalla: Writing – review \& editing. Dimiter L. Balabanski: Writing – review \& editing. Ani Aprahamian: Writing – review \& editing. Viatcheslav Sharyy: Writing – review \& editing. Dominique Yvon: Writing – review \& editing. Satoshi N. Nakamura: Writing – review \& editing, Resources. Hayk Elbakyan: Investigation, Validation, Theory, Writing – review \& editing.

\bibliographystyle{elsarticle-num}
\bibliography{biblio}

\end{document}